\documentclass[aps,prl,a4paper,twocolumn]{revtex4}

\usepackage{graphics}
\usepackage{amssymb}
\usepackage{amsmath}
\usepackage{hyperref}
\usepackage{xcolor}
\usepackage{physics}
\usepackage{graphicx}
\usepackage{fancyhdr}
\usepackage{bm}
\usepackage{braket}
\usepackage{svg}

\usepackage{times}
\usepackage{courier}
\usepackage{bm}
\usepackage{subfig}
\usepackage{mdframed}
\usepackage{dsfont}
\usepackage{tikz}
\usetikzlibrary{shapes,arrows}
\usepackage{bbm}
\usepackage{verbatim}
\usepackage{indentfirst}
\hypersetup{
    colorlinks=true,
    linkcolor=blue,
    urlcolor=cyan,
    citecolor=cyan,}

\usetikzlibrary{decorations.markings}

\newcommand{\id}{\mathds{1}}

\begin{document}


\title{Categorical-Symmetry Resolved Entanglement in CFT} 

\author{P. Saura-Bastida$^{(1)}$}
\email{pablo.saura@upct.es}

\author{A. Das$^{(2), (3)}$}
\email{arpit.das@ed.ac.uk}

\author{G. Sierra$^{(4)}$}
\email{german.sierra@csic.es}

\author{J. Molina-Vilaplana$^{(1)}$}
\email{javi.molina@upct.es}

\affiliation{${}^{(1)}$ Universidad Polit\'ecnica de Cartagena. 
Cartagena, Spain.}
\affiliation{${}^{(2)}$School of Maths, University of Edinburgh, Edinburgh, EH9 3FD, U.K.}
\affiliation{${}^{(3)}$Higgs Centre for Theoretical Physics, University of Edinburgh, Edinburgh EH8 9YL, U.K.}
\affiliation{${}^{(4)}$Instituto de Física Teórica, UAM/CSIC, Universidad Autónoma de Madrid, Madrid, Spain}

\date{\today}

\begin{abstract}
We propose a symmetry resolution of entanglement for categorical non-invertible symmetries (CaT-SREE) in $(1+1)$-dimensional CFTs. The definition parallels that of group-like invertible symmetries, employing the concept of symmetric boundary states with respect to a categorical symmetry. Our examination extends to rational CFTs, where the behavior of CaT-SREE mirrors that of group-like invertible symmetries. We find that CaT-SREE can be defined if there is no obstruction to gauging the categorical symmetry, as happens in the case of group-like symmetries. We also provide instances of the breakdown of entanglement equipartition at the next-to-leading order in the cutoff expansion. Our findings shed light on how the interplay between conformal boundary conditions and categorical symmetries lead to specific patterns in the entanglement entropy.
\end{abstract}

\maketitle

\noindent\textit{Introduction}\textemdash 
The notion of a global symmetry in quantum field theory (QFT) has been recently generalized in ways that go beyond those described by groups. 
Central to this, is the idea that every symmetry can be associated to a topological operator \cite{Gaiotto:2014}. 
The most striking of these generalizations are \emph{higher-form} symmetries, related to the conservation of extended objects, and \emph{categorical} or non-invertible symmetries, symmetries whose associated topological operators form a \emph{fusion category}, that is, do not fuse according to a simple group law.
The study of these non-invertible symmetries is providing new and deep insights into the characterization of universal properties of quantum systems of wide interest, spanning condensed matter and high-energy physics
(see \cite{Gomes:2023ahz, McGreevy:2022} for a comprehensive and pedagogical review of these developments.).

In $(1+1)$-dimensional  conformal field theories (CFT), on which we focus on this work, non-invertible symmetries implement dualities such as the Kramers-Wannier duality of the 1+1d Ising model \cite{Kramers:1941, Frohlich:2004} and the duality between momentum and winding modes (T-duality) of the free compactified boson \cite{Fuchs:2007, Thorngren:2021, Chang:2020}. In these theories, a finite categorical symmetry is defined through a fusion category ${\cal C}$ of 1-dimensional topological defect line operators (TDLs). In rational conformal field theories (RCFTs) with a {\itshape{diagonal}} modular invariant partition function \cite{Anderson:88, Schellekens:96}, these TDLs are known as Verlinde lines. Verlinde lines represent both invertible as well as non-invertible symmetries  \cite{Chang:2018, Hegde:2021sdm}. If the set of lines is denoted as $\{\mathcal{L}_i\}_{i\in\mathcal{V}}$ where $\mathcal{V}$ labels the operators $\mathcal{L}_i \in \mathcal{C}$,  the  fusion algebra is given by,
\begin{equation} 
\label{eq:fusion_lines}
    \mathcal{L}_i \times \mathcal{L}_j = \sum_{k\in\mathcal{V}} N_{ij}^{k}\,  \mathcal{L}_k \,,
\end{equation}
where $N_{ij}^{k} \in \mathbb{Z}_{\geq 0}$ are non-negative integer-valued fusion coefficients. Topological defect lines and particularly Verlinde lines $\mathcal{L}$, do not generically have an inverse $\mathcal{L}^{-1}$ such that $\mathcal{L} \times \mathcal{L}^{-1} = \id$.

Parallel to generalizing the concept of global symmetry,  there has been a remarkable interest in understanding the relation between entanglement in QFT and symmetries. In systems with a global group-like invertible symmetry, this has been carried out through the \emph{Symmetry Resolved Entanglement Entropy} (SREE) \cite{Goldstein:2017, Xavier:2018, Murciano:2020} which intuitively quantifies the amount of entanglement for different charge sectors. Remarkably, it has been shown that at leading order in the UV cutoff expansion, the SRE entropies are equal for all the charge sectors, a result known as \emph{entanglement equipartition} \cite{Xavier:2018}.

Hitherto, the entanglement behaviour in the presence of categorical symmetries in a QFT has been unknown. In this Letter, we establish the SREE for categorical symmetries (CaT-SREE), mirroring the case of group-like invertible symmetries once the notion of symmetric boundary states with respect to a categorical symmetry is provided \cite{Choi:2023}. Our results shed light on how certain CFT boundary conditions  preserve or enhance certain categorical symmetries, leading to specific patterns in the entanglement entropy. We illustrate the proposal with two RCFTs, the critical Ising model, where it is not possible to obtain a CaT-SREE and the tricritical Ising model, where it is possible, and the result at leading order shows \emph{entanglement equipartition}.

\noindent\textit{Symmetry Resolved Entanglement in CFT}\textemdash 
In extended quantum systems, the entanglement entropy (EE) measures the amount of quantum correlations between the degrees of freedom located within an arbitrary region $A$ and those sited on its complement $B$. Assuming that the Hilbert space $\mathcal{H}$  of the  system factorizes as $\mathcal{H} = \mathcal{H}_A \otimes \mathcal{H}_{B}$, where $\mathcal{H}_A$ contains the degrees of freedom in the region $A$ and $\mathcal{H}_{B}$ the ones in $B$,  for given a pure state $|\Psi\rangle \in \mathcal{H}$, the reduced density matrix of $A$ is defined by tracing out the degrees of freedom  corresponding to the complementary region $B$ as $\rho_A = \Tr_{\mathcal{H}_B} \ket{\Psi}\bra{\Psi}$.

The entanglement between $A$ and $B$ is thus quantified through the R\'enyi and entanglement entropies 
\begin{align}
\label{eq:renyi}
    S_A^n &= \frac{1}{1-n}\, \log\, \Tr\, \rho_A^n\, ,\\ \nonumber
    S_{A} &= \lim_{n \to 1}\, S_A^n = -\Tr \rho_A \log \rho_A\, .
\end{align}

We consider now there is a local charge operator $\mathcal{Q}=\mathcal{Q}_A \otimes \id_{B} + \id_{A}\otimes \mathcal{Q}_B$ 
that generates a global Abelian symmetry group $G$ in our theory. 
When $\ket{\Psi}$ is an eigenstate of $\mathcal{Q}$, then $[\rho_A, \mathcal{Q}_A]=0$ and $\rho_A$ is block-diagonal $\rho_A = \oplus_{Q}\, \Pi_Q\, \rho_A =\oplus_{Q} p_A[Q]\, \rho_A[Q]$, with $\sum_Q p_A[Q] =1$ and $\Tr \rho_A[Q]=1$, each block corresponding to a charge sector of $\mathcal{Q}_A$ where $Q$ are eigenvalues of $\mathcal{Q}_A$, $\Pi_Q$ is a projector to the eigenspace of $Q$ and $p_A[Q]=\Tr \left[\Pi_Q\, \rho_A\right]$ is the probability of measuring the charge value $Q$ in the region $A$.  $G$ being Abelian, the eigenvalues $Q$ label the irreducible  representations $r$ of the group.

As a result, the entanglement between regions $A$ and $B$, may be decomposed into the contributions of each charge sector \cite{Goldstein:2017, Xavier:2018, Murciano:2020}  through the symmetry resolved R\'enyi entropy
\begin{align}
    S^n_A[Q]=\frac{1}{1-n}\, \log\, \Tr \rho^n_A[Q]\, .
\end{align}

\emph{Entanglement equipartition} is the situation for which $\Tr \rho^n_A[Q]$ and thus $S_A[Q]$, do not depend on $Q$.  With this, the fundamental object to compute the SREE is the replica partition function \cite{Holzhey:1994, Calabrese:2004} at a fixed value of charge $Q$
\begin{align}
\label{eq:ee_q}
    Z_n[Q] = \Tr \, \Pi_Q\, \rho_A^n\, ,
\end{align}
from which the SREE can be written as
\begin{align}
    S^n_A[Q]=\frac{1}{1-n}\, \log\, \frac{Z_n[Q]}{Z[Q]^n}\, , \quad Z[Q]\equiv Z_1[Q]\, .
\end{align}

In a $(1+1)$-dimensional CFT, the factorization of the Hilbert space $\mathcal{H}$  as $\mathcal{H} = \mathcal{H}_A \otimes \mathcal{H}_{B}$, requires imposing  boundary conditions $a$, $b$ that preserve conformal symmetry at the entangling surface $\partial A$. These boundary conditions have non-trivial consequences for the EE \cite{Ohmori:2014,Northe:2023}. Specifying the region $A$ to an interval of length $\ell$, this is implemented by encircling the two entangling points at $\partial A$ with two disks of radius $\varepsilon \ll 1$, acting as UV cutoffs at which the boundary conditions $a$ and $b$ are imposed (Fig 1. upper panel).
This manifold is mapped  into an annulus of length $W= 2 \log\left(\ell/\varepsilon\right) + \mathcal{O}(\varepsilon)$ and circumference $2\pi$ ($2\pi n$, after replicating) by a conformal transformation, (down panel in Fig 1.) where the space-time is periodic in one direction and the $\ket{a}$ and $\ket{b}$ states are defined at the $\varepsilon$ boundaries.
\begin{figure}
\centering
\includegraphics[width=0.45\textwidth]{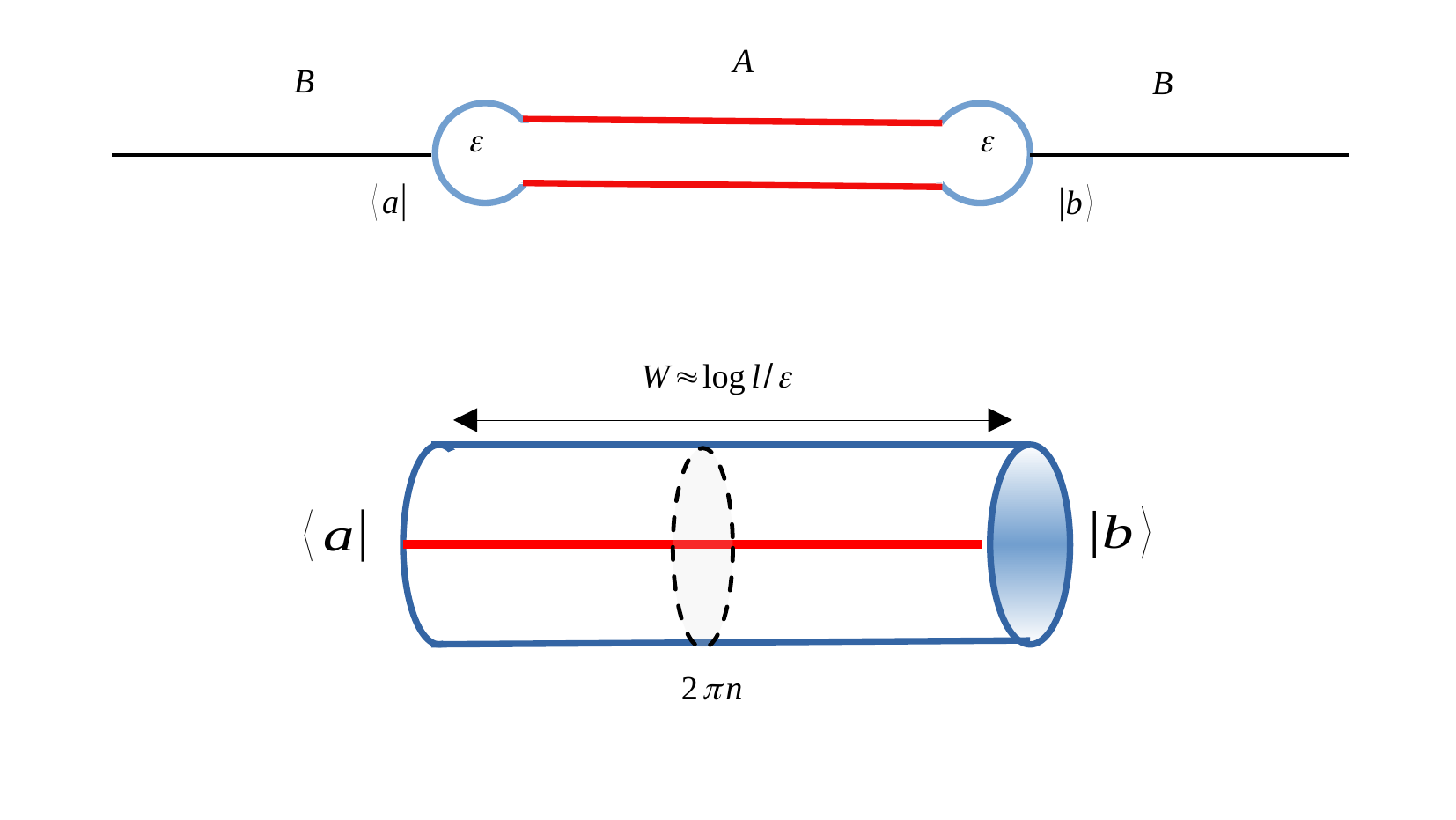}
\caption{\emph{\small The factorization $ab$ imposes disks $\varepsilon \ll 1$ with boundary conditions $a$, $b$ (upper panel). The resulting manifold is replicated and after tracing over $\mathcal{H}_{B,ba}$, a conformal transformation yields an annulus of width $W$ and circumference $2\pi n$}}
\end{figure}
In this geometry, traces of $\rho_A^n$ are evaluated in terms of BCFT partition functions as \cite{Cardy:1989,Cardy:2016}
\begin{align}
   Z_n[q^n]&= \Tr_{ab}\, [\rho_A^n] = \frac{Z_{ab}[q^n]}{Z_{ab}[q]^n}\, ,\\ \nonumber 
   Z_{ab}[q]&=\Tr_{ab} \rho_A = \Tr_{ab} \left[q^{\left(L_0 - c/24\right)}\right]\, ,
\end{align}
 with the Virasoro zero mode $L_0$ and the central charge $c$. Here, $\Tr_{ab} \equiv \Tr_{\mathcal{H}_A,ab}$ refers to a trace taking into account the non-trivial boundary conditions and $q=e^{2\pi i \tau}$ is the nome with the modular parameter $\tau = i\pi/W$. Therefore, $q=e^{-2 \pi^2/ W}$ and $\tilde{q}=e^{-2 W}$, with $\tilde{q}$ obtained after a modular S-transformation, $S:\tau \to \tilde{\tau} = -1/\tau$.

After imposing the Hilbert space decomposition $\mathcal{H}=\mathcal{H}_{A, ab} \otimes \mathcal{H}_{B,ba}$,  the remaining symmetry algebra in a CFT with a global symmetry is called $\mathcal{A}$ and $\mathcal{H}_{A, ab}  = \bigoplus_i \mathcal{H}_i^{n^i_{ab}}$
with $i$ running over the allowed representations of $\mathcal{A}$ and  the multiplicities $n^i_{ab}$ depending on the  boundary conditions $a$ and $b$. Then, the replica BCFT partition functions can be written in terms of the characters $\chi_i(q)=\Tr_{\mathcal{H}_i}\left[q^{\left(L_0 - c/24\right)}\right]$ for the representation $i$,
\begin{align}
\label{eq:bcft_amplitude}
    Z_{ab}[q^n]=\sum_i\, n^i_{ab}\, \chi_i(q^n)=\bra{a}\tilde{q}^{\frac{1}{n}\left(L_0 - c/24\right)}\ket{b}\, .
\end{align}
The last equality is obtained after a modular transformation to the $S$-dual channel where the boundary condition dependence explicitly appears in terms of Cardy conformal boundary states 
\begin{equation} \label{CardyConstruction}
\ket{a} = \sum_{j} \frac{S_{aj}}{\sqrt{S_{0j}}} \ket{j}\rangle\, ,
\end{equation}
 with $ \ket{j}\rangle$, being an Ishibashi state for the $j$-th representation of $\mathcal{A}$  \cite{Ishibahi:00},
and coefficients $S_{aj}$ are elements of the the modular matrix $\mathcal{S}$ of the CFT \cite{Cardy:1989, Verlinde:1988}. 

\textit{SREE for Group Symmetries}\textemdash 
The symmetry resolution of entanglement entropy of Abelian group symmetries has been well-studied previously \cite{Casini:2019, Magan:2021, DiGiulio:2022, Murciano:2023}. The projectors into different irreducible representations of a finite group $G$ are given by,
\begin{align} 
\label{eq:proj_Pi_r}
\Pi^r = \frac{d_r}{|G|} \sum_{g \in G} \chi^*_r(g)\,   \widehat{\mathcal{L}}_g  = \frac{d_r}{|G|} \sum_{g \in G} \chi^*_r(g)\, 
\begin{tikzpicture}
\begin{scope}[very thick,decoration={markings, mark=at position 0.5 with {\arrow{>}}}] 
    \draw[postaction={decorate}] (-0.75,-0.1)--(0.75,-0.1) node[pos=0.4,above] {$\widehat{\mathcal{L}}_g$};
\end{scope}
\node[draw=none] at (1.2,0) {$,$};
\end{tikzpicture} 
\end{align}
where $r$ labels the irreps of $G$ and thus the different $Q$-charge sectors, $d_r$ is the dimension of the irrep, $|G|$ is the order of the group, $\chi_r^*(g)$ is the character of the element of $g \in G$ in the irrep $r$ and $\mathcal{L}_g$ is the topological operator implementing the action of $g$ on states supported on the region $A$.  

Using projectors \eqref{eq:proj_Pi_r} one may write the partition function associated to a charge sector labeled by $r$ in Eq. \eqref{eq:ee_q} as:
\begin{align} \label{eq:part_func}
Z_{ab}[q^n,r] &= \Tr_{ab} \, \Pi^r \, \rho_A^n =\frac{d_r}{|G|} \sum_{g \in G} \chi^*_r(g) \frac{Z_{ab}[q^n,g]}{Z_{ab}^n[q]}\, ,\\ \nonumber
Z_{ab}[q^n,g] &=  \Tr\left[\widehat{\mathcal{L}}_g\, q^{n\left(L_0-c/24\right)}\right]\, ,
\end{align}
where the explicit action of the topological operator $\widehat{\mathcal{L}}_g$ is encoded in the \emph{charged moment} $Z_{ab}[q^n,g]$ (one for each element of the group). Here, we use $\mathcal{L}_g$ for a topological line in Euclidean spacetime, and $\widehat {\mathcal {L}}_g$ for the corresponding operator acting on the Hilbert space. As before, one may express $Z_{ab}[q^n,g]$ in the $S$-dual channel in terms of  boundary states $\ket{a}$ and $\ket{b}$ as
\begin{equation}
\label{eq:charged_Z_bcft}
Z_{ab}[q^n,g] =   {}_g{\langle a|} \tilde{q}^{\frac{1}{n}\left(L_0 - c/24\right)} |b\rangle_g\, ,
\end{equation}
where the sub-index $g$ represents that the states belong to the Hilbert space generated by inserting the operator $\widehat{\mathcal{L}}_g$ as a defect operator in the original theory, that is the \emph{defect} or \emph{twisted} Hilbert space $\mathcal{H}_{\mathcal{L}_g}$. Thus, in this approach, computing SREE reduces to find suitable boundary states $\ket{a}_g$ and $\ket{b}_g$. 
Namely, as $Z_{ab}[q^n,g]$ is defined through the insertion of $\mathcal{L}_g$ in the annulus partition function, it is required that $\mathcal{L}_g$ can end topologically on the boundary of the interval which imposes a constraint on the allowed boundary states in the dual $S$-channel \cite{Choi:2023}. 

For invertible group symmetries the topological endability is  equivalent to having $G$ invariant boundary states. A natural definition in the $S$-dual channel for a (conformal) boundary  $a$ to be $G$-symmetric is 
\begin{equation}
   \widehat{ \mathcal{L}}_h\ket{a}_g= \ket{a}_g\,,~~~\forall ~h\in G\,.
\end{equation}

For finite groups the result at leading order in the limit when $\varepsilon \ll \ell$ (where $q \to 1$ and $\tilde{q} \to 0$) is quite simple. There, the main contribution comes from the untwisted sector \cite{Inthisfootnote}, that is to say, the vacuum
state propagation is the major contribution to the amplitude in the $S$-dual channel and the SREE reads as \cite{Casini:2019, Magan:2021, DiGiulio:2022, Murciano:2023}
\begin{align} 
\label{FiniteGroupResult}
S_A[q,r] = \frac{c}{6}\, \log \frac{\ell}{\varepsilon} + \log \frac{d_r^2}{|G|} +  {\rm g}_a + {\rm g}_b\, ,
\end{align}
where   ${\rm g}_a=\log \braket{0|a}$, ${\rm g}_b=\log \braket{0|b}$ are the Affleck-Ludwig  boundary entropies \cite{Affleck:1991},  and  $d_r$ is the dimension of the irrep $r$. 
The term $\mathcal{O}(\log \ell/\varepsilon)$ captures the equipartition of EE among distinct charge sectors, primarily at the leading order. This equal distribution is broken by the term of order $\mathcal{O}((\ell/\varepsilon)^0)$ by the negative term $\log p_r$, with $p_r = d^2_r/|G|$ representing the probability of measuring the representation $r$ within block $A$, a scenario denoted as weak entanglement equipartition to distinguish it from the strong equipartition. A parallel outcome was observed in the examination of WZW models \cite{Calabrese:2021}.

\textit{Categorical-Symmetry Resolved Entanglement Entropy}\textemdash 
We propose the symmetry resolution of entanglement for categorical non-invertible symmetries (CaT-SREE) in analogy with the BCFT approach for group-like invertible symmetries. For this, it is necessary to define topological endability and thus, symmetric boundary conditions, for the case of (categorical) non-invertible  symmetries. These have been proposed in  \cite{Choi:2023} through the notions of \emph{strongly symmetric} and \emph{weakly symmetric}  boundary states. While these two concepts are equivalent for invertible group-like symmetries, they diverge for category-like non-invertible symmetries.

Recalling the fusion algebra in Eq.\eqref{eq:fusion_lines},  we focus on the finite subset of boundary conditions  $\{a \}_{a\in\mathcal{B}}$ with $\mathcal{B}$ labeling these boundaries, related by the action of a finite symmetry fusion category $\mathcal{C}$. The corresponding boundary states are denoted as $\{\ket{a}\}$. This is known as a \emph{module category} and the action of $\mathcal{L}_i$ acting on such a class of boundary $a$ is given by
\begin{equation}
    \mathcal{L}_i \otimes a = \bigoplus_{b\in\mathcal{B}} \widetilde{N}_{ia}^{b}\,  b \,,
\end{equation}
where $\widetilde{N}_{ia}^{b} \in \mathbb{Z}_{\geq 0}$.  With this, two notions of $\mathcal{C}$-symmetric boundary states can be established \cite{Choi:2023}:
A conformal  boundary condition $a$ is $\mathcal{C}$-\emph{strongly symmetric} if the corresponding boundary state $\ket{a}$ is an eigenstate under the action of $\mathcal{C}$ with eigenvalues given by the quantum dimensions $\braket{\mathcal{L}_i}$,
\begin{equation} \label{eq:strong}
    \widehat{\mathcal{L}}_i \ket{a} = \braket{\mathcal{L}_i}\ket{a} \quad \forall \mathcal{L}_i\in \mathcal{C}\, .
\end{equation}
This definition reduces to a $G$-symmetric boundary condition in the case of group-like invertible symmetries. On the other hand it is considered that a conformal boundary condition $a$ is $\mathcal{C}$-\emph{weakly symmetric} if every topological line in $\mathcal{C}$ can end topologically on $a$. Operationally speaking this means that $\widetilde{N}_{ia}^a \ge1$ for every $\mathcal{L}_i$ in $\mathcal{C}$, which implies
    \begin{equation}
    \label{eq:weakly}
        \widehat{\mathcal{L}}_i \ket{a} = \ket{a} \oplus \cdots \quad \forall \mathcal{L}_i\in \mathcal{C}\, .
    \end{equation}
This second notion of $\mathcal{C}$-symmetric boundary condition relax enough the requirements for finding the appropriate boundary states needed to define SREE for fusion categorical non-invertible symmetries.

\textit{CaT-SREE in RCFT}\textemdash 
The simplest models to define the SREE for fusion categorical symmetries (CaT-SREE) are two-dimensional rational CFTs (RCFT), for which there exists a correspondence between Verlinde lines and bulk primary operators \cite{InRCFTfootnote}. Thus, each line representing a symmetry of the model is associated with one primary operator, and their fusion rules are those given  by the operator product expansion (OPE) coefficients of the corresponding primaries.

 The first step to define CaT-SREE is to write a full set of projectors associated to the elements of fusion category $\mathcal{C}$ in terms of elements of the modular matrix $\mathcal{S}$ of the CFT \cite{Lin:2022}:
\begin{align} \label{FullProj}
\begin{split}
\begin{tikzpicture}
\node[draw=none] at (-0.75,0) {$\Pi_a{}^c = \sum S_{0c} \bar{S}_{bc} \quad$};
\node[draw=none] at (-0.925,-0.375) {${}_b$};
\begin{scope}[very thick] 
    \draw[postaction={decorate}] (0.8,0)--(1.3,0);
    \draw[postaction={decorate}] (1.5,-0.35)--(1.5,0);    
\end{scope}
\begin{scope}[very thick,decoration={markings, mark=at position 0.5 with {\arrow{>}}}] 
    \draw[postaction={decorate}] (1.5,-0.25)--(1.5,0.75) node[pos=0.65,left] {$\widehat{\mathcal{L}}_a$};
    \draw[postaction={decorate}] (1.7,0)--(3.0,0) node[pos=0.65,above] {$\widehat{\mathcal{L}}_b$};
\end{scope}
\node[draw=none] at (3.5,0) {$,$};
\end{tikzpicture} 
\end{split}
\end{align}
where $\{\widehat{\mathcal{L}}_a\}_{a \in \mathcal{V}} \in \mathcal{C}$. The lines pictorially represent the Verlinde lines of the RCFT  inserted either along the time direction in the annulus (vertical ones), which twist the Hilbert space of the theory, or along the spatial direction, which amounts to charged operators acting over the states on the  Hilbert  space (horizontal ones) \cite{Chang:2018}. 
 
As we are interested in resolving EE on the original Hilbert space of the theory, we will consider only projectors of the form $\Pi_{\id}^c$. Here, we write these projectors in full analogy with the group-like symmetry case Eq. \eqref{eq:proj_Pi_r} as:
\begin{equation} \label{Proj}
\Pi_\id{}^c := \Pi^c = \frac{d_c}{|\mathcal{C}|} \sum_{b \in \mathcal{C}} \chi^*_c(b)\quad
\begin{tikzpicture}
\begin{scope}[very thick,decoration={markings, mark=at position 0.5 with {\arrow{>}}}] 
    \draw[postaction={decorate}] (-1.0,-0.25)--(1.0,-0.25) node[pos=0.5,above] {$\widehat{\mathcal{L}}_b$};
\end{scope}
\node[draw=none] at (1.2,0) {$,$};
\end{tikzpicture} 
\end{equation}
by defining  $d_c = \frac{S_{0c}}{S_{00}}$ as the quantum dimension of the line $\widehat{\mathcal{L}}_c$, the order of the category $|\mathcal{C}| = \sum_c d^2_c$, and the characters $\chi^*_c(b) = \frac{\bar{S}_{bc}}{S_{00}}$. 

We note that the projectors in Eq.\eqref{Proj} are written for a simple element of the category $\mathcal{L}_c\in \mathcal{C}$. However, the element labeling an irrep of the category is not, in general, a simple object and may be described by non-simple topological lines whose associated projectors can be written as \eqref{Proj}  \cite{Bhardwaj1:2023, Bhardwaj2:2023}.

Thus, in analogy with group-like invertible symmetries,  we define the CaT-SREE in terms of the partition functions
\begin{equation}
\label{eq:cat_Z}
     Z_{c_1 c_2}[q^n,a] = \Tr_{c_1 c_2}\left[\Pi{}^a \, \rho_A^n\right] =\frac{d_a}{|\mathcal{C}|} \sum_{b \in \mathcal{C}} \chi^*_a(b) \frac{Z_{c_1 c_2}[q^n,b]}{\left(Z_{c_1 c_2}[q]\right)^n}\, ,
\end{equation}
where the  generalized charged moment in the $S$-dual channel is defined as 
\begin{equation}
     Z_{c_1 c_2}[q^n,b] =   {}_b{\langle c_1|} \tilde{q}^{\frac{1}{n}\left(L_0 - c/24\right)} |c_2\rangle_b\, ,
\end{equation}
and $\ket{c_{1,\,2}}_b$ are  Cardy boundary states in the $\mathcal{C}$-weakly symmetric sense exposed above.

We illustrate our definition with two examples, the critical Ising model and the tricritical Ising model.

\textit{The Critical Ising Model}\textemdash
The critical Ising model is described by a  $(1+1)$-dimensional RCFT with a central charge $c=\frac{1}{2}$. 
There are three primary operators in the model: the identity $\id$, the energy field $\epsilon$ and the spin field $\sigma$. 
The symmetries of this model are described by three Verlinde lines: $\{\widehat{\id},\widehat{\eta}\}$ which conform the usual $\mathbb{Z}_2$ symmetry of the Ising model, and  $\widehat{\mathcal{N}}$ that implements the Kramers-Wannier duality \cite{Kramers:1941, Frohlich:2004}. 
These lines follow the fusion rules of the Ising category
\begin{align}
    \eta \times \eta = \id\, ,\quad 
    \mathcal{N} \times \mathcal{N} = \id + \eta\, , \quad
    \eta \times \mathcal{N}=\mathcal{N}
\end{align}

As discussed above, to obtain the CaT-SREE one must first compute  the set of the $\mathcal{C}_{\rm Ising}$-symmetric Cardy states  through \eqref{CardyConstruction}. In doing so, it is noticed that there are three simple boundary states in this model, 
\begin{align}
\label{eq:bc_ising}
\begin{split}
\ket{\id} & = \frac{1}{\sqrt{2}} \ket{\id}\rangle + \frac{1}{\sqrt{2}} \ket{\epsilon}\rangle + \frac{1}{2^{1/4}} \ket{\sigma}\rangle \, ,
\\
\ket{\epsilon} & = \frac{1}{\sqrt{2}} \ket{\id}\rangle + \frac{1}{\sqrt{2}} \ket{\epsilon}\rangle - \frac{1}{2^{1/4}} \ket{\sigma}\rangle \, ,
\\
\ket{\sigma} & = \ket{\id}\rangle - \ket{\epsilon}\rangle \, .
\end{split}
\end{align}

The boundary states $\ket{\id}$, $\ket{\epsilon}$, and $\ket{\sigma}$ conform the Ising, or more technically, the Tambara-Yamagami $\mathrm{TY}_+\mathbb{Z}_2$ fusion category, that is a regular module category. Therefore, there is a one-to-one correspondence between the  boundary states \eqref{eq:bc_ising} and the Verlinde lines such that  $\ket{\id}\equiv \hat{\mathds{1}},\ket{\epsilon} \equiv \hat{\eta}, \ket{\sigma}\equiv \widehat{\mathcal{N}}$ with 
\begin{align}
\label{eq:VerlindeIsing_bc}
\begin{split}
   & \hat{\eta} \ket{\id} =\ket{\epsilon}\,, \quad  \hat{\eta}\ket{\epsilon} = \ket{\id}\,, \quad \hat{\eta}\ket{\sigma} = \ket{\sigma}\,, \\ 
   & \widehat{\mathcal{N}}\ket{\id} = \ket{\sigma}\,, \quad \widehat{\mathcal{N}}\ket{\epsilon}  = \ket{\sigma}\,, \quad \widehat{\mathcal{N}}\ket{\sigma} = \ket{\id}\oplus\ket{\epsilon}\,.
\end{split}
\end{align}
We note that only $\ket{\sigma}$ is invariant under the action of the group $\mathbb{Z}_2$. In this sense, $\ket{\sigma}$ is a $\mathbb{Z}_2$-symmetric state, and thus it is possible to use it to compute the SREE for the $\mathbb{Z}_2$ group-like symmetry of the model \cite{Murciano:2023}. 
 However, none of the three boundary states are symmetric, neither in the strong nor weak sense, under the action of $\mathcal{N}$. As a result, it is not possible to define the CaT-SREE in the critical Ising model. A fusion category can only admit a strongly symmetric boundary if it is anomaly-free, while it admits a weakly symmetric boundary if and only if it can be "gauged" in the generalized sense posed in \cite{Choi:2023}. Being $\mathcal{C}_{\rm Ising}$ anomalous, hence the impossibility of defining the CaT-SREE from  the obstruction to gauging it, as just happens in the case of group-like symmetries \cite{Li:2022, Murciano:2023}.

\textit{The Tricritical Ising Model}\textemdash
The tricritical Ising model is a RCFT with central charge $c=\frac{7}{10}$. The model is composed by six primary operators and six lines. In addition to the trivial line $\mathds{1}$ and the $\mathbb{Z}_2$ invertible line $\eta$, there are four more simple lines, $W$, $\eta W$, $\mathcal{N}$, and $W\mathcal{N}$. Non-trivial fusion rules for these lines are given by:
\begin{align}
& \eta \times \eta = \mathds{1} \: , \quad \mathcal{N} \times \mathcal{N} = \mathds{1} + \eta \, ,\\ \nonumber
& \eta \times \mathcal{N} = \mathcal{N} \times \eta = \mathcal{N} \: , \quad 
W \times W=\mathds{1} + W \quad .
\end{align}
From these relations we can identify $\{\mathds{1},\eta,\mathcal{N}\}$ as a TY${}_+\mathbb{Z}_2$ subcategory and $\{\mathds{1},W\}$ a Fibonacci subcategory $\mathcal{C}_{\rm Fib}$. Same as with critical Ising model, for the first group of lines we cannot find symmetric boundary conditions, neither strong nor weak. However, that is not the case for $\mathcal{C}_{\rm Fib}$ that is the simplest example of a category that can be gauged, and therefore admits a weakly symmetric boundary. Namely, there are three boundary states that are weakly symmetric under $\mathcal{C}_{\rm Fib}$:
\begin{align}
&\widehat{W} |W\rangle =  |W\rangle \oplus |\mathds{1}\rangle \: , \quad \widehat{W} |\eta W\rangle =  |\eta W\rangle \oplus |\eta\rangle \: , \\ \nonumber
&\widehat{W} |W \mathcal{N}\rangle =  |W \mathcal{N} \rangle \oplus |\mathcal{N}\rangle \quad .
\end{align}

For simplicity, we choose to work with the boundary condition $\ket{W\mathcal{N}}$. Through Eq. \eqref{CardyConstruction}, this state can be written as: 
\begin{equation}
|W \mathcal{N} \rangle =   \frac{1}{\sqrt{N}} \left(|\mathds{1}\rangle\rangle + \varphi^{-3/2} |\epsilon \rangle \rangle - \varphi^{-3/2} |\epsilon^\prime \rangle \rangle - |\epsilon^{\prime \prime} \rangle \rangle\right) \, ,
\end{equation}

with $\varphi = \frac{1+\sqrt{5}}{2}$ the golden ratio and $N=\left( \frac{10}{5 + 2\sqrt{5}} \right)^{1/2}$. Thus, the charged moment associated to the untwisted sector is given in terms of the Virasoro characters by see Supplementary Material:
\begin{align}
Z_{W \mathcal{N}}[q^n, \mathds{1}] &= \frac{1}{N} \Bigg[\chi_0\left(\tilde{q}^{\frac{1}{n}}\right)+ \varphi^{-3} \chi_{\frac{1}{10}}\left(\tilde{q}^{\frac{1}{n}}\right) \\ \nonumber
&+ \varphi^{-3} \chi_{\frac{3}{5}}\left(\tilde{q}^{\frac{1}{n}}\right) + \chi_{\frac{3}{2}}\left(\tilde{q}^{\frac{1}{n}}\right)\Bigg] \, ,
\end{align}
where, for notational convenience we use $Z_{aa}\to Z_{a}$.

In order to compute the charged moment $Z_{W \mathcal{N}}[q^n, W]$, we need an explicit expression of the boundary state $\ket{W \mathcal{N}}$ twisted by the introduction of the Verlinde line $\widehat{W}$ as a defect operator. In general, this new boundary state is given by a combination of twisted Ishibashi states, that is, conformal scalars on the $W$-twisted Hilbert space. The twisted  $W$-Hilbert space contains 9 primary operators; among them there are 3 scalars, $\epsilon_{W},\, \epsilon^\prime_{W},\, \sigma_{W}$ with conformal weights $\frac{1}{10}$, $\frac{3}{5}$ and $\frac{3}{80}$ respectively. This implies that the $\mathcal{C}_{\rm Fib}$-symmetric twisted Cardy state is a linear combination of the twisted Ishibashi states associated with these operators:
\begin{equation}
|W \mathcal{N}\rangle_W = \alpha_1 |\epsilon\rangle\rangle_W + \alpha_2 |\epsilon^\prime\rangle\rangle_W + \alpha_3 |\sigma\rangle\rangle_W \, ,
\end{equation}
for some fixed coefficients $\alpha_i$. With this state, the twisted charged moment is
\begin{equation}
Z_{W \mathcal{N}}[q^n, W] = \alpha_1^2  \chi_{\frac{1}{10}}\left(\tilde{q}^{\frac{1}{n}}\right) + \alpha_2^2 \chi_{\frac{3}{5}}\left(\tilde{q}^{\frac{1}{n}}\right) + \alpha_3^2 \chi_{\frac{3}{80}}\left(\tilde{q}^{\frac{1}{n}}\right) \, .
\end{equation}
Comparing the two contributions for the CaT-SREE of the $\mathcal{C}_{\rm Fib}$ subcategory of this model, one notices that $Z_{W \mathcal{N}}[q^n, \mathds{1}]$ dominates over $Z_{W \mathcal{N}}[q^n, W]$ at leading order in the $\varepsilon$-expansion (see Supplementary Material). 

In order to find the CaT-SREE, we note that there are 2 irreps of $\mathcal{C}_{\rm Fib}$ that we label with $\mathrm{r}_{\mathcal{C}}=\{A,B\}$. With this, one may write the projectors associated to those representations as a combination of projectors of the form \eqref{Proj},
\begin{align}
   \Pi^A &= \Pi^\id + \Pi^\eta + \Pi^{\mathcal{N}}\, \\ \nonumber
    \Pi^B &= \Pi^W + \Pi^{\eta W} + \Pi^{W\mathcal{N}} \, ,
\end{align}
that explicitly read as
\begin{equation} \label{eq:tricrit-proj}
\Pi^{\mathrm{r}_{\mathcal{C}}}  = \frac{d_{\mathrm{r}_{\mathcal{C}}}}{|\mathcal{C}|}\left(d_{\mathrm{r}_{\mathcal{C}}}\, 
\begin{tikzpicture}
\begin{scope}[very thick,decoration={markings, mark=at position 0.5 with {\arrow{>}}}] 
    \draw[postaction={decorate}] (-0.75,0)--(0.75,0) node[pos=0.65,above] {$\hat{\id}$};
\end{scope}
\end{tikzpicture} +\,  \chi^*_{\mathrm{r}_{\mathcal{C}}}(W)\, 
\begin{tikzpicture}
\begin{scope}[very thick,decoration={markings, mark=at position 0.5 with {\arrow{>}}}] 
    \draw[postaction={decorate}] (-0.75,0)--(0.75,0) node[pos=0.65,above] {$\widehat{W}$};
\end{scope}
\end{tikzpicture} 
\, \right) \, ,
\end{equation}
where $d_A=1$ and $d_B=\varphi$ are the quantum dimensions of the Fibonacci anyons, the total quantum dimension is given by the usual formula $|\mathcal{C}| = d_A^2 + d_B^2 = 1 + \varphi^2$, and the characters are $\chi_A^*(W) = \varphi$ and $\chi_B^*(W) = -1$. 

These projectors coincide with those characterizing a theory with Fibonacci anyons, those describing the low-energy physics of the 
fractional quantum Hall effect at filling factor $\nu = 5/2 $ \cite{Feiguin:2006}.

As there are two simple anyons in this category, there are two projectors of the type of Eq. \eqref{Proj}. This ensures that the projectors \eqref{eq:tricrit-proj} successfully project into the irreps of the Fibonacci subcategory $\mathcal{C}_{\rm Fib}$.

With this, the partition functions associated to each charged sector \eqref{eq:part_func} are given by:
\begin{equation}
Z[q^n,\mathrm{r}_{\mathcal{C}}] =  \frac{d_\mathrm{r_{\mathcal{C}}}}{|\mathcal{C}|} \left( d_\mathrm{r_{\mathcal{C}}}\, \frac{Z[q^n,\mathds{1}]}{(Z^n[q]} + \chi^*_{\mathrm{r_{\mathcal{C}}}}(W)\frac{Z[q^n,W]}{Z^n[q])} \right) \, ,
\end{equation}
where the subscript $W\mathcal{N}$ has been suppressed for convenience. From these, the CaT-SREE at leading order for both irreps $\mathrm{r}_{\mathcal{C}}=\{A,B\}$ reads as
\begin{align}
\begin{split}
S[q,\mathrm{r}_{\mathcal{C}}] = \frac{c}{3} \log\frac{\ell}{\varepsilon} + \log \frac{d_{\mathrm{r}_{\mathcal{C}}}^2}{|\mathcal{C}|} + 2\, {\rm g}_{W \mathcal{N}} \, ,
\end{split}
\end{align}
with ${\rm g}_{W\mathcal{N}}=\log \langle\braket{\id|W\mathcal{N}} $ the Affleck-Ludwig boundary entropy.
This is the main result in this work. Formally, it is analogous to the one obtained for finite groups \eqref{FiniteGroupResult}, that is, the CaT-SREE at leading order in the UV cutoff, is equally distributed among the different (Fibonacci anyon) charge sectors $A$ and $B$. Similarly to invertible group-like symmetries, the entanglement equipartition is broken by constant terms related to the quantum dimension $d_{\rm r_{{\mathcal{C}}}}$ and the boundary entropy.

We note that the same entanglement resolution can be obtained for the Tetracritical Ising model, whose topological lines generate the non-anomalous symmetry  ${\rm Rep}(S_3)$ which posses weakly symmetric boundary states under the subcategory $\mathcal{C}_{\rm Fib}$. Finally, we remark that in some cases, the CaT-SREE can be obtained in terms of strongly symmetric states, as, for instance, the double Ising CFT, which admits a strongly symmetric boundary condition with respect to the categorical symmetry ${\rm Rep}(H_8)$  \cite{Choi:2023}.
\\

\noindent\textit{Conclusions}\textemdash 
We have shown how categorical symmetries shape the entanglement structure in CFTs by proposing a symmetry resolution of entanglement for these symmetries (CaT-SREE). Our proposal has then provided new insights into the relationship between EE and boundary conditions. As the EE of a topological phase reflects the fusion rules of anyons and their braiding statistics, a deeper understanding of the imprint of categorical symmetries on entanglement provides a future route to characterize topological order. It is also interesting to understand in future works, the microscropic description of CaT-SREE by exploring the entanglement properties of non-invertible operators on the lattice \cite{Aasen:2016dop, Aasen:2020jwb, Seiberg:2023cdc, Seiberg:2024gek}. 
\\

\noindent\textit{Acknowledgments}\textemdash
We thank Nabil Iqbal, Sridip Pal, Thomas Bartsch, Matthew Bullimore, Karan Bhatia, Tudor Dimofte, Sa\v{s}o Grozdanov, Giorgio Frangi, Joan Simon Soler, Naveen Balaji Umasankar and Mile Vrbica for insightful discussions on this project. The work of P.S.-B. is supported by Fundaci\'on S\'eneca 
de la Regi\'on de Murcia, grant 21609/FPI/21. A.D. is supported by the STFC Consolidated Grant ST/T000600/1 -- ``Particle Theory at the Higgs Centre''.  J.M.-V. thanks  Programa Recualificaci\'on del Sistema Universitario Español 2021–2023. P.S.-B. and J.M.-V. thank the financial support of Spanish Ministerio de Ciencia e Innovación PID2021-125700NAC22. G.-S. thanks the grant  PID2021-127726NB-I00.
\\

\bibliographystyle{utphys}

\begin{thebibliography}{99}

\bibitem{Gaiotto:2014}
D.~Gaiotto, A.~Kapustin, N.~Seiberg and B.~Willett,
JHEP \textbf{02} (2015), 172
doi:10.1007/JHEP02(2015)172
[arXiv:1412.5148 [hep-th]].

\bibitem{Gomes:2023ahz}
P.~R.~S.~Gomes,
SciPost Phys. Lect. Notes \textbf{74} (2023)
doi: 10.21468/SciPostPhysLectNotes.74
[arXiv:2303.01817 [hep-th]].

\bibitem{McGreevy:2022}
J.~McGreevy,
doi:10.1146/annurev-conmatphys-040721-021029
[arXiv:2204.03045 [cond-mat.str-el]].

\bibitem{Kramers:1941}
H.~A.~Krammers and G.~H.~Wannier,
Phys. Rev. \textbf{60}, (1941) 252
10.1103/PhysRev.60.252

\bibitem{Frohlich:2004}
J.~Frohlich, J.~Fuchs, I.~Runkel and C.~Schweigert,
Phys. Rev. Lett. \textbf{93} (2004), 070601
doi:10.1103/PhysRevLett.93.070601
[arXiv:cond-mat/0404051 [cond-mat]].


\bibitem{Fuchs:2007}
J.~Fuchs, M.~R.~Gaberdiel, I.~Runkel and C.~Schweigert,
J. Phys. A \textbf{40} (2007), 11403
doi:10.1088/1751-8113/40/37/016
[arXiv:0705.3129 [hep-th]].

\bibitem{Thorngren:2021}
R.~Thorngren and Y.~Wang,
[arXiv:2106.12577 [hep-th]]

\bibitem{Chang:2020}
C.~M.~Chang and Y.~H.~Lin,
JHEP \textbf{10} (2021), 125
doi:10.1007/JHEP10(2021)125
[arXiv:2012.01429 [hep-th]].


\bibitem{Anderson:88}
G.~Anderson and G.~Moore,
Commun. Math. Phys. \textbf{117}, 441–450 (1988)
doi:10.1007/BF01223375 


\bibitem{Schellekens:96}
A.~N.~Schellekens,
Fortschritte der Physik \textbf{44}, 605-705 (1996)
doi:10.1002/prop.2190440802


\bibitem{Chang:2018}
C.~M.~Chang, Y.~H.~Lin, S.~H.~Shao, Y.~Wang and X.~Yin,
JHEP \textbf{01} (2019), 026
doi:10.1007/JHEP01(2019)026
[arXiv:1802.04445 [hep-th]].


\bibitem{Hegde:2021sdm}
S.~Hegde and D.~P.~Jatkar,
Mod. Phys. Lett. A \textbf{37} (2022) 29
doi:10.1142/S0217732322501930
[arXiv:2101.12189 [hep-th]].


\bibitem{Goldstein:2017}
M.~Goldstein and E.~Sela,
Phys. Rev. Lett. \textbf{120} (2018) no.20, 200602
doi:10.1103/PhysRevLett.120.200602
[arXiv:1711.09418 [cond-mat.stat-mech]].


\bibitem{Murciano:2020}
S.~Murciano, G.~Di Giulio and P.~Calabrese,
JHEP \textbf{08} (2020), 073
doi:10.1007/JHEP08(2020)073
[arXiv:2006.09069 [hep-th]].


\bibitem{Xavier:2018}
J. C. Xavier, F. C. Alcaraz, and G. Sierra, 
Phys. Rev. B {\bf  98}, 041106 (2018).
doi:10.1103/PhysRevB.98.041106

\bibitem{Choi:2023}
Y.~Choi, B.~C.~Rayhaun, Y.~Sanghavi and S.~H.~Shao,
Phys. Rev. D \textbf{108} (2023) no.12, 125005
doi:10.1103/PhysRevD.108.125005
[arXiv:2305.09713 [hep-th]].


\bibitem{Holzhey:1994}
C.~Holzhey, F.~Larsen and F.~Wilczek,
Nucl. Phys. B \textbf{424} (1994), 443-467
doi:10.1016/0550-3213(94)90402-2
[arXiv:hep-th/9403108 [hep-th]].

\bibitem{Calabrese:2004}
P.~Calabrese and J.~L.~Cardy,
J. Stat. Mech. \textbf{0406} (2004), P06002
doi:10.1088/1742-5468/2004/06/P06002
[arXiv:hep-th/0405152 [hep-th]].


\bibitem{Ohmori:2014}
K.~Ohmori and Y.~Tachikawa,
J. Stat. Mech. \textbf{1504} (2015), P04010
doi:10.1088/1742-5468/2015/04/P04010
[arXiv:1406.4167 [hep-th]].

\bibitem{Northe:2023}
C.~Northe,
Phys. Rev. Lett. \textbf{131} (2023) no.15, 151601
doi:10.1103/PhysRevLett.131.151601
[arXiv:2303.07724 [hep-th]].

\bibitem{Cardy:1989}
J.~L.~Cardy,
Nucl. Phys. B \textbf{324} (1989), 581-596
doi:10.1016/0550-3213(89)90521-X

\bibitem{Cardy:2016}
J.~Cardy and E.~Tonni,
J. Stat. Mech. \textbf{1612} (2016) no.12, 123103
doi:10.1088/1742-5468/2016/12/123103
[arXiv:1608.01283 [cond-mat.stat-mech]].

\bibitem{Ishibahi:00} 
Ishibashi states are those that satisfy  
$\left(L_n - \bar{L}_{-n}\right)\ket{j}\rangle=0$ with $L_n$ Virasoro modes
and  $\langle\langle j|\tilde{q}^{\left(L_0-c/24\right)}|i\rangle\rangle =\chi_i(\tilde{q})\, \delta_{ij}$.

\bibitem{Verlinde:1988}
E.~P.~Verlinde,
Nucl. Phys. B \textbf{300} (1988), 360-376
doi:10.1016/0550-3213(88)90603-7

\bibitem{Casini:2019}
H.~Casini, M.~Huerta, J.~M.~Mag\'an and D.~Pontello,
JHEP \textbf{02} (2020), 014
doi:10.1007/JHEP02(2020)014
[arXiv:1905.10487 [hep-th]].

\bibitem{Magan:2021}
J.~M.~Magan,
JHEP \textbf{12} (2021), 100
doi:10.1007/JHEP12(2021)100
[arXiv:2111.02418 [hep-th]].

\bibitem{Murciano:2023}
Y.~Kusuki, S.~Murciano, H.~Ooguri and S.~Pal,
[arXiv:2309.03287 [hep-th]].

\bibitem{DiGiulio:2022}
G.~Di Giulio, R.~Meyer, C.~Northe, H.~Scheppach and S.~Zhao,
SciPost Phys. Core \textbf{6} (2023), 049
doi:10.21468/SciPostPhysCore.6.3.049
[arXiv:2212.09767 [hep-th]].

\bibitem{Inthisfootnote}
In this language, untwisted is equivalent to twisted by the identity element $e$. That is, the untwisted Hilbert space is the original Hilbert space $\mathcal{H}_e \equiv \mathcal{H}$.

\bibitem{Affleck:1991}
I.~ Affleck,  and A.~Ludwig, 
Phys. Rev. Lett. \textbf{67},(1991)161
10.1103/PhysRevLett.67.161,

\bibitem{Calabrese:2021}
P.~Calabrese, J.~Dubail and S.~Murciano,
JHEP \textbf{10} (2021), 067
doi:10.1007/JHEP10(2021)067
[arXiv:2106.15946 [hep-th]].

\bibitem{InRCFTfootnote}
In RCFT this is called \emph{regular module category}. In these,  $\mathcal{B}=\mathcal{V}$ and  $\widetilde{N}_{ia}^{b}$  are given by the fusion coefficients of bulk primary operators which establishes a direct correspondence between the Cardy boundary conditions and the Verlinde lines \cite{Choi:2023}.

\bibitem{Lin:2022}
Y.~H.~Lin, M.~Okada, S.~Seifnashri and Y.~Tachikawa,
JHEP \textbf{03} (2023), 094
doi:10.1007/JHEP03(2023)094
[arXiv:2208.05495 [hep-th]].


\bibitem{Bhardwaj1:2023}
L.~Bhardwaj and S.~Schafer-Nameki,
[arXiv:2304.02660 [hep-th]].

\bibitem{Bhardwaj2:2023}
L.~Bhardwaj and S.~Schafer-Nameki,
[arXiv:2305.17159 [hep-th]]



\bibitem{Li:2022}
L.~Li, C.~T.~Hsieh, Y.~Yao and M.~Oshikawa,
[arXiv:2205.11190 [hep-th]].

\bibitem{Feiguin:2006}
A.~Feiguin, S.~Trebst, A.~W.~W.~Ludwig, M.~Troyer, A.~Kitaev, Z.~Wang and M.~H.~Freedman,
Phys. Rev. Lett. \textbf{98} (2007), 160409
doi:10.1103/PhysRevLett.98.160409
[arXiv:cond-mat/0612341 [cond-mat.str-el]].

%

\bibitem{Aasen:2016dop}
D.~Aasen, R.~S.~K.~Mong and P.~Fendley
J. Phys. A \textbf{49} (2016) no.35, 354001
doi:10.1088/1751-8113/49/35/354001
[arXiv:1601.07185 [cond-mat.stat-mech]].

\bibitem{Aasen:2020jwb}
D.~Aasen, P.~Fendley and R.~S.~K.~Mong
[arXiv:2008.08598 [cond-mat.stat-mech]].

\bibitem{Seiberg:2023cdc}
N.~Seiberg and S.~H.~Shao
[arXiv:2307.02534 [cond-mat.str-el]].

\bibitem{Seiberg:2024gek}
N.~Seiberg, S.~Seifnashri and S.~H.~Shao
[arXiv:2401.12281 [cond-mat.str-el]].

\bibitem{Moore:1989vd}
G.~W.~Moore and N.~Seiberg,
``Lectures on RCFT,''
1989 Banff NATO ASI: Physics, Geometry and Topology (1989)
report number: RU-89-32, YCTP-P13-89

\bibitem{MMS}
S.~D.~Mathur, S.~Mukhi and A.~Sen,
``On the Classification of Rational Conformal Field Theories,''
Phys. Lett. B \textbf{213} (1988), 303--308
doi:10.1016/0370-2693(88)91765-0

\bibitem{Beer:2018}
K.~Beer, D.~Bondarenko, A.~Hahn, M.~Kalabakov, N.~Knust, L.~Niermann, T.~J.~Osborne, C.~Schridde, S.~Seckmeyer and D.~E.~Stiegemann, \textit{et al.}
[arXiv:1811.06670 [quant-ph]].

\bibitem{Gils:2009}
C.~Gils, S.~Trebst, A.~Kitaev, Alexei, A.~Ludwig, M.~Troyer and Z.~Wang,
"Topology-driven quantum phase transitions in time-reversal-invariant anyonic quantum liquids,"
Nature Physics \textbf{5} (2009) no.11, 834-839
doi:10.1038/nphys1396
[arXiv:0906.1579 [cond-mat.str-el]].

\bibitem{Read:1998}
N.~Read and E.~Rezayi,
``Beyond paired quantum Hall states: Parafermions and incompressible states in the first excited Landau level,''
Phys. Rev. B \textbf{59} (1999), 8084
doi:10.1103/PhysRevB.59.8084
[arXiv:cond-mat/9809384 [cond-mat]].


\end{thebibliography}


\onecolumngrid
\appendix

\section{A brief review of rational CFTs and Topological Defect Lines}
A rational CFT is a $(1+1)$-dimensional CFT which has a finite set of primary fields with respect to some, possibly extended chiral algebra $\hat{\mathfrak{g}}_k$ \cite{Moore:1989vd}. It was shown in \cite{Anderson:88} that for such a CFT the central charge and the conformal dimensions of the primaries are all rational numbers. The modular invariant torus partition function of an RCFT is given as \cite{MMS}:
\begin{align}\label{torus_gen}
    Z(\tau,\Bar{\tau}) = \sum_{i,j=0}^{n-1} \, M_{ij} \, \chi_i(\tau)\Bar{\chi}_j(\Bar{\tau})
\end{align}
where $\tau$ is the modular parameter on the torus; $Z(\gamma\tau,\gamma\Bar{\tau}) = Z(\tau,\Bar{\tau})$ with $\gamma\in\text{SL}_2(\mathbb{Z})$; $\chi(\tau)$ (or, $\Bar{\chi}(\Bar{\tau})$) are holomorphic (or, anti-holomorphic) characters associated with the highest weight integral representations of the full chiral algebra $\hat{\mathfrak{g}}_k$; $n$ denotes the number of linearly independent characters in the theory and $M_{ij}$ is the multiplicity matrix which captures the number of primaries which share the same characters -- which usually happens when the Dynkin diagram of the corresponding simple Lie algebra $\mathfrak{g}_k$ has some symmetries. If in a given RCFT, $M_{ij}$ is diagonal then it is called a {\itshape{diagonal}} RCFT and the partition function becomes,
\begin{align}
    Z(\tau,\Bar{\tau})= |\chi_0|^2 + \sum_{i=1}^{n-1}Y_i|\chi_i|^2, \label{diag_part_rcft}
\end{align}

The positive integers $Y_i$ in Eq.(\ref{diag_part_rcft}) are the multiplicities of the characters, and the number of primaries, $p$, is given in terms of these by $p=1+\sum_{i=1}^{n-1}Y_i$. 

From here on we will only consider diagonal RCFTs. Let us denote the primaries by $|\phi_i\rangle$. For such theories, the action of the topological defect lines (TDLs) on the bulk Hilbert space is defined as \cite{Chang:2018, Hegde:2021sdm},
\begin{align}
    \hat{\mathcal{L}}_k |\phi_i\rangle = \frac{S_{ki}}{S_{0i}}|\phi_i\rangle. \label{TDL_act}
\end{align}
Note from the above definition, the action of a TDL preserves the conformal dimensions. In the case of diagonal RCFTs, TDLs are referred to as the Verlinde lines. It is known that a TDL, say $\hat{\mathcal{L}}_k$ corresponding to a primary $|\phi_k\rangle$ commutes with the full chiral algebra $\hat{\mathfrak{g}}_k$ \cite{Chang:2018}.

We can now define torus partition functions with Verlinde lines inserted along different non-contractible cycles of the torus \cite{Hegde:2021sdm}. TDLs along spatial directions are encountered while taking the trace to compute the partition function as operators' insertions. TDLs along temporal direction impose a boundary condition on each spatial slice and hence modify the Hilbert space over which the trace is performed. Given a Verlinde line as in (\ref{TDL_act}), we can get the partition function with a defect insertion (that is along the spatial cycle) as,
\begin{align}
    Z^{\hat{\mathcal{L}}_k} = \text{Tr}\left(\hat{\mathcal{L}}_k \, e^{-q\left(L_0 - \frac{c}{24}\right)}\right) = \sum\limits_{i}\frac{S_{ki}}{S_{0i}}\chi_i(\tau)\Bar{\chi}_i(\Bar{\tau}), \label{Z_sp}
\end{align}
From the above, we can obtain the defect partition function along the temporal cycle by performing an $S$-transformation: $S:\tau\to -\frac{1}{\tau}$,
\begin{align}
    Z_{\hat{\mathcal{L}}_i}(\tau,\Bar{\tau}) = Z^{\hat{\mathcal{L}}_i}\left(-\frac{1}{\tau},-\frac{1}{\Bar{\tau}}\right) = \sum\limits_{l,m,k} \frac{S_{ik}S_{lk}\bar{S}_{mk}}{S_{0k}}\chi_l\Bar{\chi}_m = \sum\limits_{l,m} N^m_{il} \chi_l\Bar{\chi}_m, \label{sp_t}
\end{align}
where $N^m_{il}$ are the fusion coefficients (which are non-negative integers) appearing in the fusion algebra of the primaries,
\begin{align}
    [i]\times[j] = \sum\limits_k N^{k}_{ij}[k]. \label{fusion_prim}
\end{align}
These coefficients give the degeneracy of the operators in the defect Hilbert space with conformal dimensions $(h,\Bar{h})$. Thus, we can read off the operator content in the defect Hilbert space by expressing $Z_{\hat{\mathcal{L}}}$ in terms of the characters. Note, the Verlinde lines share the same fusion algebra as that of the primaries.

\section{Leading  contribution for SREE for finite groups}
Following \cite{Ohmori:2014}, let us consider $\ket{0}$ to be the state with smallest scaling dimension $\Delta_0$ in a concrete sector with non zero overlap to both boundary states $\ket{a}$ and $\ket{b}$.  In the limit $\varepsilon \ll \ell$, the leading contribution to $ Z_{ab}[q^n]=\bra{a}\tilde{q}^{\frac{1}{n}\left(L_0 - c/24\right)}\ket{b}\,$ can be evaluated as 
\begin{equation}
\label{eq:Zqe}
	Z_{ab}[q^n]
	\sim \braket{a|0} e^{-\frac{2 W}{n}\left(\Delta_0-\frac{c}{24}\right)} \braket{0|b}\, ,
\end{equation} 
When $\ket{0}$ is the vacuum, $\Delta_0=0$. Therefore, when computing the EE using the result above, the leading term is $\sim W$ plus a constant term given by the quantities $g_a = \log \braket{a|0}$ and $g_b = \log \braket{b|0}$ known as the Affleck-Ludwig boundary entropies \cite{Affleck:1991}. 

Regarding subleading terms, we denote $\ket{\Delta_{\mathcal{O}}}$ the state with the second-lowest scaling dimension $\Delta_{\mathcal{O}}> \Delta_0$ with nonzero overlap to both boundary states $\ket{a}$ and $\ket{b}$. With this we have
\begin{align}
  Z_{ab}[q^n]&= 
	\braket{a|0} e^{\frac{W}{n}\frac{c}{12}} \braket{0|b} + \braket{a|\Delta_{\mathcal{O}}}e^{\frac{W}{n}\left(\frac{c}{12}-\Delta_{\mathcal{O}}\right)} \braket{\Delta_{\mathcal{O}}|b} +\cdots.
\end{align} 
That is, the next to leading corrections to R\'enyi entropies are of the order  $O\left(e^{-\Delta_{\mathcal{O}}\, W/n}\right)$. As $\mathcal{O}$  must have non-zero overlap with both boundary conditions, therefore the subleading correction must depend on the choice of the boundary conditions.

Now we use the results above for the SREE. The leading contributions for the charged moment $Z_{ab}[q^n,e]$ is given by Eq. \eqref{eq:Zqe}. Regarding the charge moment
\begin{align}
Z_{ab}[q^n,g] =   {}_g{\langle a|} \tilde{q}^{\frac{1}{n}\left(L_0 - c/24\right)} |b\rangle_g\, ,
\end{align}

as commented in the main text, the  boundary states $\ket{a,b}_g$ in this amplitude pertain to a the Hilbert space $\mathcal{H}_{\mathcal{L}_g}$. The ground state  of this twisted Hilbert space has scaling dimension $\Delta^{(g)}_0 >\Delta^{(e)}_0 = \Delta_0 $, that is, greater than that of  the untwisted sector. As a consequence, in the in $\tilde{q}\to 0$ limit we have
     \begin{align}
     \label{eq:leading_twisted}
       Z_{ab}[q^n,g]&\equiv   {}_g \bra{a} \tilde{q}^{\frac{1}{n}(L_0-\frac{c}{24})}   \ket{b} {}_g \sim {}_g \braket{a|\Delta^{(g)}_0}e^{-\frac{2 W}{n}\left(\Delta^{g}_0-\frac{c}{24}\right)}\braket{\Delta^{g}_0|b}{}_g \, .
     \end{align}
Therefore, one sees that the leading contribution from the untwisted sector $Z_{ab}[q^n,e]$ dominates over the leading contribution to $Z_{ab}[q^n,g]$ in the computation of $S_A[q,r]$, thus recovering Eq. (13) in the main text.

\section{The Tricritical Ising Model}
The tricritical Ising model is a RCFT with central charge $c=\frac{7}{10}$. As a particularity of a minimal model, most quantities are completely determined in terms of the modular $\mathcal{S}$-matrix. In this model, the $\mathcal{S}$-matrix is given by:

\begin{equation}
    \mathcal{S}= \frac{1}{\sqrt{5}}
    \begin{pmatrix}
        s_2 & s_1 & s_1 & s_2 & \sqrt{2} s_1 & \sqrt{2} s_2\\
        s_1 & -s_2 & -s_2 & s_1 & \sqrt{2} s_2 & -\sqrt{2} s_1\\
        s_1 & -s_2 & -s_2 & s_1 & -\sqrt{2} s_2 & \sqrt{2} s_1\\
        s_2 & s_1 & s_1 & s_2 & -\sqrt{2} s_1 & -\sqrt{2} s_2\\
        \sqrt{2} s_1 & \sqrt{2} s_2 & -\sqrt{2} s_2 & -\sqrt{2} s_1 & 0 & 0\\
        \sqrt{2} s_2 & -\sqrt{2} s_1 & \sqrt{2} s_1 & -\sqrt{2} s_2 & 0 & 0\\
    \end{pmatrix} \, .
\end{equation}

\noindent Where $s_1 = \sin(2\pi/5)$ and $s_2=\sin(4\pi/5)$. The symmetries of the tricritical Ising model are described by a fusion category with six TDLs. The action of the topological operators implementing the symmetries can be computed through the formula:

\begin{equation}
    \hat{\mathcal{L}}_i \ket{\phi_j} =  \frac{S_{ij}}{S_{0j}} \ket{\phi_j} \, .
\end{equation}

\noindent Being $\ket{\phi_j}$ the state corresponding to the primary operator $\phi_j$ through the operator-state map. The action of this lines is summarized in Table \ref{table:lines-tric}. 

\begin{table}
\begin{center}
\begin{tabular}{cccccccc}
 & & $\mathds{1}$ & $\epsilon$ & $\epsilon^\prime$ & $\epsilon^{\prime \prime}$ & $\sigma$ & $\sigma^\prime$
\\
$\widehat{\mathds{1}}$ : & \quad & 1 & 1 & 1 & 1 & 1 & 1
\\
$\widehat{\eta W}$ : & $\quad$ & $\varphi$ & $-\varphi^{-1}$ & $-\varphi^{-1}$ & $\varphi$ & $\varphi^{-1}$ & $-\varphi$
\\
$\widehat{W}$ : & \quad & $\varphi$ & $-\varphi^{-1}$ & $-\varphi^{-1}$ & $\varphi$ & $-\varphi^{-1}$ & $\varphi$
\\
$\widehat{\eta}$ : & \quad & 1 & 1 & 1 & 1 & -1 & -1
\\
$\widehat{W \mathcal{N}}$ : & \quad & $\sqrt{2}\varphi$ & $\sqrt{2}\varphi^{-1}$ & $- \sqrt{2} \varphi^{-1}$ & $-\sqrt{2} \varphi$ & 0 & 0
\\
$\widehat{\mathcal{N}}$ : & \quad & $\sqrt{2}$ & $-\sqrt{2}$ & $\sqrt{2}$ & $-\sqrt{2}$ & 0 & 0
\end{tabular}
\end{center}
\caption{\textit{Verlinde lines in the tricritical Ising model.  The model has six primary bulk fields: the identity, three thermal and two spin fields. $\varphi \equiv \frac{s_1}{s_2} = \frac{1+\sqrt{5}}{2}$ is the golden ratio.}}
\label{table:lines-tric}
\end{table}

Being a RCFT, in this model,  simple boundary states are fully determined by the $\mathcal{S}$-matrix. The expansion of the boundary states in terms of the Ishibashi states is known as the Cardy construction:

\begin{equation}
    \ket{c_i} = \sum_{j}\frac{S_{ij}}{\sqrt{S_{0j}}} \ket{j}\rangle \, .
\end{equation}

Through the Cardy construction, one finds three $\mathcal{C}$-weakly symmetric boundary states for $\mathcal{C}_{\rm Fib}$, that is

\begin{equation}
    \hat{W} \ket{W} = \ket{W} \oplus \ket{\id} \, , \quad  \hat{W} \ket{\eta W} = \ket{\eta W} \oplus \ket{\eta} \, , \quad \hat{W} \ket{W \mathcal{N}} = \ket{W \mathcal{N}} \oplus \ket{\mathcal{N}} \, ,
\end{equation}

which are explicitly given by,

\begin{align}
\begin{split}
    \ket{W} & = \frac{1}{\sqrt{N_{W}}} \left(\ket{\id}\rangle - \varphi^{-3/2} \ket{\epsilon}\rangle - \varphi^{-3/2} \ket{\epsilon^\prime}\rangle + \ket{\epsilon^{\prime \prime}}\rangle - 2^{1/4} \varphi^{-3/2} \ket{\sigma}\rangle + 2^{1/4}\ket{\sigma^\prime}\rangle \right) \, , 
    \\
    \ket{\eta W} & = \frac{1}{\sqrt{N_{\eta W}}} \left(\ket{\id}\rangle - \varphi^{-3/2} \ket{\epsilon}\rangle - \varphi^{-3/2} \ket{\epsilon^\prime}\rangle + \ket{\epsilon^{\prime \prime}}\rangle + 2^{1/4} \varphi^{-3/2} \ket{\sigma}\rangle - 2^{1/4}\ket{\sigma^\prime}\rangle \right) \, , 
    \\
    \ket{W \mathcal{N}} & = \frac{1}{\sqrt{N_{W\mathcal{N}}}} \left(\ket{\id}\rangle + \varphi^{-3/2} \ket{\epsilon}\rangle - \varphi^{-3/2} \ket{\epsilon^\prime}\rangle - \ket{\epsilon^{\prime \prime} \rangle }\right)\, , 
\end{split}
\label{eq:cardy_boundaries}
\end{align}

 with

\begin{equation}
  N_{W} = \left(\frac{20}{5+2\sqrt{5}}\right)^{1/2} \,  ,   \quad N_{\eta W} = \left(\frac{20}{5+2\sqrt{5}}\right)^{1/2} \, ,  \quad N_{W\mathcal{N}} = \left(\frac{10}{5+2\sqrt{5}}\right)^{1/2} \, .
\end{equation}

We would like to emphasize that the Ishibashi states are a very convenient choice of basis for the the boundary Cardy states, as they form an orthonormal basis satisfying:
\begin{equation}
\langle \bra{i} \tilde{q}^{\frac{1}{n}(L_0 - c/24)} \ket{j} \rangle = \chi_i\left(\tilde{q}^{\frac{1}{n}}\right) \delta_{ij} \, ,
\end{equation}
\noindent being $\chi_i(\tilde{q}^{\frac{1}{n}})$ the $i$-th Virasoro character. One may  easily realize that this is a very useful property in order to compute the charged moments. Taking, for instance, the state $\ket{W \mathcal{N}}$, the identity charged moment in Eq. (27) of the main text is obtained as:
\begin{align}
\begin{split}
Z_{W\mathcal{N}}[q^n,\id] = & \bra{W \mathcal{N}} \tilde{q}^{\frac{1}{n}(L_0 - c/24)} \ket{W \mathcal{N}} 
\\
= & \frac{1}{N_{W \mathcal{N}}} \left[ \langle \bra{\id} \tilde{q}^{\frac{1}{n}(L_0 - c/24)} \ket{\id} \rangle + \varphi^{-3} \langle \bra{\epsilon} \tilde{q}^{\frac{1}{n}(L_0 - c/24)} \ket{\epsilon} \rangle \right.
\\
+ & \left. \varphi^{-3} \langle \bra{\epsilon^\prime} \tilde{q}^{\frac{1}{n}(L_0 - c/24)} \ket{\epsilon^\prime} \rangle + \langle \bra{\epsilon^{\prime \prime}} \tilde{q}^{\frac{1}{n}(L_0 - c/24)} \ket{\epsilon^{\prime \prime}} \rangle \right]
\\
= &  \frac{1}{N_{W \mathcal{N}}} \left[ \chi_0\left(\tilde{q}^{\frac{1}{n}}\right) + \varphi^{-3} \chi_{\frac{1}{10}}\left(\tilde{q}^{\frac{1}{n}}\right) + \varphi^{-3} \chi_{\frac{3}{5}}\left(\tilde{q}^{\frac{1}{n}}\right)+ \chi_{\frac{3}{2}}\left(\tilde{q}^{\frac{1}{n}}\right) \right] \, ,
\end{split}
\end{align}
\noindent where we have used the orthogonality property to drop cross terms directly.

From the Cardy states in \eqref{eq:cardy_boundaries}, one may compute the boundary states in the $W$-twisted sector of the theory. In general, these twisted boundary states are given by a combination of twisted Ishibashi states, that is, conformal scalars on the $W$-twisted Hilbert space. The twisted  $W$-Hilbert space contains 9 primary operators; among them there are 3 scalars, $\epsilon_{W},\, \epsilon^\prime_{W},\, \sigma_{W}$ with conformal weights $\frac{1}{10}$, $\frac{3}{5}$ and $\frac{3}{80}$ respectively. This implies that the $\mathcal{C}_{\rm Fib}$-symmetric twisted Cardy states $|c_i\rangle_W$ must be a linear combination of the twisted Ishibashi states associated to these operators, that is:

\begin{equation}
|c_i\rangle_W = \alpha_{1,\,i} |\epsilon\rangle\rangle_W + \alpha_{2,\,i} |\epsilon^\prime\rangle\rangle_W + \alpha_{3,\,i} |\sigma\rangle\rangle_W \, ,
\end{equation}

with some fixed coefficients for each twisted Ishibashi state. As the $\ket{\id}\rangle$ Ishibashi state is not present in any twisted Cardy state $|c_i\rangle_W$,  using the argument given in the previous section, one concludes that the untwisted sector is dominant in the calculation of SREE at leading order. Therefore, the only contribution coming from imposing different boundary conditions will be that of the Affleck-Ludwig boundary entropy, which in general will be of the form:

\begin{equation}
    {\rm g}_i = \log \langle \braket{\id|c_i} = \log \frac{1}{\sqrt{N_i}} \, ,
\end{equation}

for each state imposed as a boundary condition.

\section{Fibonacci Anyons}
Anyons arise as the low-energy quasiparticle excitations in many-body systems with topological order. They are observed in the fractional quantum Hall effect and their exotic properties can be experimentally characterized by measuring scattering amplitudes. We refer the reader to \cite{Beer:2018} for a modern exposition of these paticles in terms of categories.

In this context, Fibonacci anyons \cite{Gils:2009,Read:1998} describe the low-energy physics of the fractional quantum Hall effect at filling factor $\nu=\frac{5}{2}$, and are relevant in topological quantum computation as they allow for universal quantum computation. In the $\mathcal{C}_{\rm Fib}$, the simple objects are the \emph{anyonic charges} denoted by $\{\id, \tau\}$.  The category $\mathcal{C}_{\rm Fib}$ is defined in terms of the fusion rules
\begin{align}
	\id\times\id&= \id\\ \nonumber
	\id\times \tau &= \tau\times\id=\tau\\ \nonumber
	\tau\times\tau &= \id+\tau
\end{align}

\noindent where $\id$ is the identity operator. Some properties of the category are:
\begin{enumerate}
\item  All objects in this category can be expressed as a sum of simple objects, e.g,
						\begin{align}
							\tau\times\tau\times\tau= \id+ 2\tau.
						\end{align}
\item The quantum dimensions are given by $d_\id=1$ and $d_\tau=\varphi$ with $\varphi = \frac{1+\sqrt{5}}{2}$ the golden ratio.

\item The $\mathcal{S}$-matrix is given by:
\begin{equation}
    \mathcal{S} = \frac{1}{\sqrt{1 + \varphi^2}} \begin{pmatrix}
        1 & \varphi \\
        \varphi & -1 \\
    \end{pmatrix} \, .
\end{equation}
which is invertible.
\end{enumerate}

 Because of the fusion rules of the anyons we can see that the global symmetry of this theory is described by the same category as the Fibonacci subcategory in the tricritical Ising model. Morover, if we build the projectors into irreps using Eq. (18) of the main text, we find:

\begin{align}
\begin{split}
\Pi^\id  & = \frac{d_\id}{d_\id^2 + d_\tau^2}\left(d_\id\, 
\begin{tikzpicture}
\begin{scope}[very thick,decoration={markings, mark=at position 0.5 with {\arrow{>}}}] 
    \draw[postaction={decorate}] (-0.75,0)--(0.75,0) node[pos=0.65,above] {$\widehat{\id}$};
\end{scope}
\end{tikzpicture} +\,  \chi^*_\id(\tau)\, 
\begin{tikzpicture}
\begin{scope}[very thick,decoration={markings, mark=at position 0.5 with {\arrow{>}}}] 
    \draw[postaction={decorate}] (-0.75,0)--(0.75,0) node[pos=0.65,above] {$\widehat{\tau}$};
\end{scope}
\end{tikzpicture} 
\, \right) \, ,
\\
\Pi^\tau & = \frac{d_\tau}{d_\id^2 + d_\tau^2}\left(d_\tau\, 
\begin{tikzpicture}
\begin{scope}[very thick,decoration={markings, mark=at position 0.5 with {\arrow{>}}}] 
    \draw[postaction={decorate}] (-0.75,0)--(0.75,0) node[pos=0.65,above] {$\widehat{\id}$};
\end{scope}
\end{tikzpicture} +\,  \chi^*_\tau(\tau) \, 
\begin{tikzpicture}
\begin{scope}[very thick,decoration={markings, mark=at position 0.5 with {\arrow{>}}}] 
    \draw[postaction={decorate}] (-0.75,0)--(0.75,0) node[pos=0.65,above] {$\widehat{\tau}$};
\end{scope}
\end{tikzpicture} 
\,  \right) \, ,
\end{split}
\end{align}

\noindent with $\chi^*_\id(\tau) = \varphi$ and $\chi^*_\tau(\tau) = -1$. We can identify these expressions as the projectors for the Fibonacci subcategory of the tricritical Ising model given in Eq. (31) of the main text.

\end{document}